# Elucidating the high compliance mechanism by which the urinary bladder fills under low pressures

Fatemeh Azari[1], Anne M. Robertson[1,2,*], Yasutaka Tobe[1], Paul N. Watton[1,3], Lori A. Birder[4,5], Naoki Yoshimura[5,6], Kanako Matsuoka[6], Christopher Hardin[7], and Simon Watkins[8]

[1]Department of Mechanical Engineering and Materials Science, University of Pittsburgh, PA, U.S.A.
[2]Department of Bioengineering, University of Pittsburgh, PA, U.S.A.
[3]Department of Computer Science & Insigneo Institute for in silico Medicine, University of Sheffield, Sheffield, U.K.
[4]Department of Medicine, University of Pittsburgh, Pittsburgh, PA, U.S.A.
[5]Department of Pharmacology and Chemical Biology, University of Pittsburgh, Pittsburgh, PA, U.S.A.
[6]Department of Urology, University of Pittsburgh, Pittsburgh, PA, U.S.A.
[7]Department of Nutrition and Exercise Physiology, University of Missouri School of Medicine, Columbia, MO, U.S.A.
[8]Center for Biologic Imaging, University of Pittsburgh, Pittsburgh, PA, U.S.A.
*Corresponding author: rbertson@pitt.edu

## ABSTRACT

The high compliance of the urinary bladder during filling is essential for its proper function, enabling it to accommodate significant volumetric increases with minimal rise in transmural pressure. This study aimed to elucidate the physical mechanisms underlying this phenomenon by analyzing the ex vivo filling process in rat from a fully voided state to complete distension, without preconditioning, using three complementary imaging modalities. High-resolution micro-CT at 10.8 $\mu$m resolution was used to generate detailed 3D reconstructions of the bladder lumen, revealing a 62 fold increase in bladder volume during filling. Pressure-volume studies of whole bladder delineated three mechanical filling regimes: an initial high-compliance phase, a transitional phase, and a final high-pressure phase. While prior studies conjectured small mucosal rugae (∼450 $\mu$m) are responsible for the high compliance phase, multiphoton microscopy (MPM) of the dome of the voided bladder revealed large folds an order of magnitude larger than these rugae. Bladder imaging during the inflation process demonstrated flattening of these large scale folds is responsible for volume increases in the initial high compliance phase. The 3D reconstructions of the bladder lumen in the filled and voided state revealed a high voiding efficiency of 97.13% ± 2.42%. The MPM imaging results suggest the large scale folds in the dome enable this high voiding fraction by driving urine toward the bladder outlet. These insights are vital for computational models of bladder biomechanics and understanding changes to bladder function due to pathological conditions such as bladder outlet obstruction and age-related dysfunction.

## Introduction

High compliance during filling is one of the hallmarks of a healthy bladder and indicates the ability of the bladder to fill with small increases in pressure. Loss of compliance can arise from a variety of medical conditions including Bladder Outlet Obstruction (BOO),[1,2] spinal cord injury[3], and cancer[4]. A reduction of bladder compliance has important clinical implications including urinary incontinence, reduced bladder capacity, increased bladder pressure and bladder wall thickening (fibrosis)[5,6]. Clinical measurements of bladder compliance are used in the diagnosis and management of conditions such as neurogenic bladder, interstitial cystitis, and bladder outlet obstruction, where targeted interventions are necessary to preserve bladder and kidney function[6,7].

From a clinical perspective, bladder compliance is a single parameter defined in the 2002 report of the International Continence Society (ICS) as the ratio of the change in bladder volume ($\Delta V$) to the change in detrusor pressure ($\Delta P$) during bladder filling, expressed in milliliters per centimeter of $H_2O$[8]. The changes are calculated between two clinically relevant points: (1) the initiation of filling, typically, when the bladder is fully voided, and (2) the point of maximum cystometric capacity, or just prior to the onset of a detrusor contraction that may precipitate leakage[8].

The gold standard approach for obtaining this data is a pressure flow urodynamic study in which vesical pressure, $P_{\text{ves}}$, is measured from a pressure sensor placed in the bladder using a catheter and $P_{\text{abd}}$ is measured using a second sensor in the rectum, vagina, or stoma, as applicable. The detrusor pressure $P_{det}$ is defined as the difference between the the vesical pressure, $P_{ves}$, and the abdominal pressure, $P_{abd}$. The clinical value of compliance is then calculated from the resulting curve for $P_{det}$ as a function of infused volume curve. More specifically, it is the inverse of slope of a line connecting the two clinically relevant points defined above.

Bladder compliance is influenced by the mechanics of the entire organ during filling - including its geometry and the mechanical properties of the wall. The bladder wall is a highly heterogeneous organ, both in terms of its wall thickness and wall components. Its biomechanical properties are a function of its microstructure, comprising multiple layers that synergistically contribute to its function under varying physiological conditions[9–12]. The innermost layer, the urothelium, is a highly specialized epithelial lining that functions as a barrier against the toxic components of urine, while also serving a neurologic function[13,14]. It consists of superficial umbrella cells, intermediate cells, and basal cells[15,16]. The umbrella cells, characterized by their tight junctions and a thick glycosaminoglycan layer, confer impermeability and protect the underlying tissues from urinary toxins[9,13]. Beneath the urothelium lies the lamina propria, a connective tissue layer rich in vasculature and innervation within an extracellular matrix (ECM). This ECM comprises collagen (types I and III) and elastin fibers, organized in a complex network[10,17]. The LP layer is surrounded by the detrusor smooth muscle layer, the primary muscular layer of the bladder, consisting of smooth muscle bundles, and collagen fibers[9,15]. The contractility of the detrusor smooth muscle is regulated by autonomic nervous system inputs, predominantly parasympathetic nerves releasing acetylcholine to stimulate muscle contraction during voiding.

During filling from the voided state, all layers of the bladder must undergo large elastic deformations sufficient to enable the volume to increase over 10-fold, depending on the species. For example, the healthy human male bladder was found to increase in volume by 28.4-fold (mean age: 24 years, $n = 28$), in pressure-flow studies conducted on normal adult volunteers[18]. During filling, each layer must accommodate this large increase in size. One mechanism for extensibility of the inner two layers of the bladder wall are macroscopic rugae, approximately 100 microns in size. These undulations have been shown to gradually flatten under increasing strain with minimal resistance[19–21]. Undulations of collagen fibers, on the order of tens of microns, provide another mechanism for extensibility. As a result of these undulations, collagen fibers can accommodate large extensions before straightening sufficiently to bear load. In addition, earlier studies have proposed that the high extensibility of the bladder's outer wall is facilitated by the structural organization of smooth muscle cell (SMC) bundles interconnected by wavy collagen fibers in an unloaded state[20]. The reorientation and elongation of these SMC bundles, along with the straightening of the wavy collagen fibers, are suggested as mechanisms for extensibility of the detrusor smooth muscle (DSM) layer. While these studies provide evidence of multiple mechanisms for compliance, to date, there has been no direct study demonstrating these mechanisms are sufficient to explain the large change in volume and associated high compliance that are collectively crucial for bladder function. We hypothesize there are additional mechanisms involved in enabling high compliance during filling from the voided state. Of importance, while these earlier compliance studies considered the extensibility of unloaded bladder tissue, they did not consider the extension of the voided (contracted) bladder to the unloaded state.

The main objective of the present study is to determine the physical mechanisms underlying low-pressure bladder filling from the voided state. We conducted two sub-studies to achieve this goal. In the first sub-study, high-resolution micro-CT imaging was performed ex-vivo on air-filled bladders to assess changes in wall morphology from the voided to the filled state. By filling the bladder with air, both the inner and outer surfaces could be imaged. Scanning multiphoton microscopy was then used to determine the structural explanation for these morphological differences. The second sub-study aimed to correlate these morphological changes with the pressure-volume curves of the bladder. For this latter study, a custom imaging-inflation system was employed to gather data on the changing bladder morphology along the pressure-volume curve and to confirm the mechanism of low-pressure bladder filling.

## Methods

### Oversight and Protocol for Animal Management
All animal procedures adhered to institutional guidelines and received approval from the Institutional Animal Care and Use Committees of Pittsburgh University, animal facilities services. Sprague-Dawley rats were housed in groups of two male and female within an environment regulated for temperature, humidity, and pressure, maintaining a 12-hour light/dark cycle, ($n = 7$). They were given free access to standard chow and water.

### Harvest of the Urinary Bladder
Sprague Dawley rats (3-4 months old) were euthanized via $CO_2$ inhalation. Voiding was confirmed by visible evidence of urine in paper within the transport box. Postmortem, the abdominal region was disinfected with ethanol to minimize friction and ensure a sterile surface. An abdominal incision was performed using surgical scissors to access the internal cavity. The integumentary and muscular layers were incised to expose the urinary bladder. Adjacent adipose tissue and ligaments were carefully excised using tweezers and surgical scissors. To stabilize the bladder during the dissection of surrounding tissues, the upper ligament near the bladder dome was clamped with forceps and positioned around the animal's forelimbs. The pubic bone was transected using surgical scissors.

Upon successful transection of the pubic bone, the urethra was exposed. In male bladders, the genital was excised, and a segment of polyethylene tubing (PE50, Fisher Scientific, Hampton, NH, USA) was inserted into the urethra, enveloped by a muscular layer. The tubing was secured to the bladder using 3-0 sutures. Subsequently, the prostate gland was excised, and the ureters were identified and cauterized near the bladder using a Thread Burner (PEN-510.00, Eurotool, South Carolina, USA). Finally, the bladder was excised from the body.

## Post-Harvest Preparation of Voided Bladder

To prevent smooth muscle cell contraction, following harvest, the bladder was immediately immersed in a solution of Hank's Balanced Salt Solution (HBSS) with the following constituents (millimolar concentrations): NaCl 138, KCl 5, $KH_2PO_4$ 0.3, $NaHCO_3$ 4, $MgCl_2$ 1, HEPES 10, and glucose 5.6, maintaining a pH of 7.4 and an osmolarity of 310 mOsm/L, in the absence of calcium. EDTA (0.5 mM) was incorporated into this solution. Furthermore, the voltage-dependent calcium channel antagonist nifedipine (5 μM; Sigma) and the sarco/endoplasmic reticulum $Ca^{2+}$-ATPase (SERCA) inhibitor thapsigargin (1 μM; Tocris Biosciences) were also included[22]. The bladder was then positioned on a Petri dish with adhesive measuring rulers for later demarcation of the bladder's longitudinal and circumferential dimensions. High-definition images of both ventral and dorsal facets of the mounted bladder were obtained using an OLYMPUS SZX10 with a focal setting of 0.63.

## High-Resolution Micro-CT Analysis of Air-Filled Bladder Geometry

Here, we briefly detail the principal steps to develop a 3D reconstructed model of each bladder using data sets from high-resolution scanning (Skyscan 1272 scanner, Bruker Micro-CT, Kontich, Belgium), Fig. 1. In the first study (N=3,), the bladder was harvested in the voided state and inflated with air (so the lumen surface could be identified). Micro-CT data were acquired at two states: (1) the voided stated (harvested condition) and (2) after ex-vivo inflation with air. In the second study (N=3), the bladder was also harvested in the voided state and then imaged in real-time during gradual infusion with a contrast liquid to obtain pressure-volume curves.

### Preparation of Voided Bladder for Micro-CT

First the PE50 tubing was affixed to the needle with cyanoacrylate glue (Loctite 414). Then the tubing was gently threaded into the urethra and just into the base of the bladder. In preparation for micro-CT imaging of each harvested female bladder (N=3), urine was gently drawn from the bladder and replaced with air in 0.1 ml increments using a1 ml syringe. Care was taken to avoid any sign of bladder inflation. Next, cyanoacrylate glue (Loctite 414) was applied at the distal end of the urethra to secure the attachment to the mounting needle and prevent air leakage during inflation. Moisture on the outer bladder surface was delicately removed by blotting with Kim wipes to avoid artifacts during micro-CT imaging.

### Acquisition of High-Resolution Micro-CT Data in Voided Bladder

Each mounted bladder was then positioned within a custom-engineered holder to ensure immobilization and prevent motion artifacts during the scanning process, Fig. 1. The holder, with the Luer-lock adapter, was sealed using parafilm and secured in the micro-CT system. Scans were performed at an 80 kV source voltage and 125 μA source current, capturing images at a 10.8 μm pixel resolution with a 2048 x 2048 frame size, a 0.6-degree rotation increment, and an exposure duration of 400 ms. The scanning duration was maintained under 10 minutes to avoid volume changes due to dehydration.

### Micro-CT Scanning in the Filled Bladder State

Following micro-CT scanning in the voided state, the 25-gauge needle hub was connected to a syringe pump (BS-8000, Braintree Scientific Inc.). Without pre-conditioning, air was administered into the voided bladder at a rate of 1.5 ml/hour to achieve a gradual filling time on the order of 30-35 minutes. Filling was continued until the specified "filled" transmural pressure was reached, after which the valve was sealed, and the bladder was detached from the apparatus. Micro-CT scanning in the filled state was then performed using the same protocol as for the voided state. Filled pressures of 35 mmHg, 57 mmHg, and 80 mmHg were used for Bladders A, B, and C, respectively. Pressure was measured using a pressure transducer (PX409, Omega Engineering Inc.).

### Post Processing of High-Resolution Micro-CT Data

The Z stacks of 2D micro-CT images were then 3D reconstructed using NRecon software (Bruker Micro-CT, Kontich, Belgium), Fig.1. The following settings were used: smoothing at level 1, ring artifact correction at 50%, and 2% beam hardening correction. Morphological analysis of the 3D model was conducted with grayscale thresholding to create masks using Simpleware ScanIP software (Synopsys, Sunnyvale, California). Segmentation of the internal (lumen) and external (ablumen) geometries was executed using Meshmixer software (Autodesk, San Francisco, California), followed by thickness analysis utilizing Materialise 3-matics software (Materialise GmbH, Munich, Germany)[23]. Wall thickness and associated parameters were determined from the finalized STL files using the midplane thickness tool in Materialise 3-matics. Associated statistics such as median, average, and standard deviation were obtained.

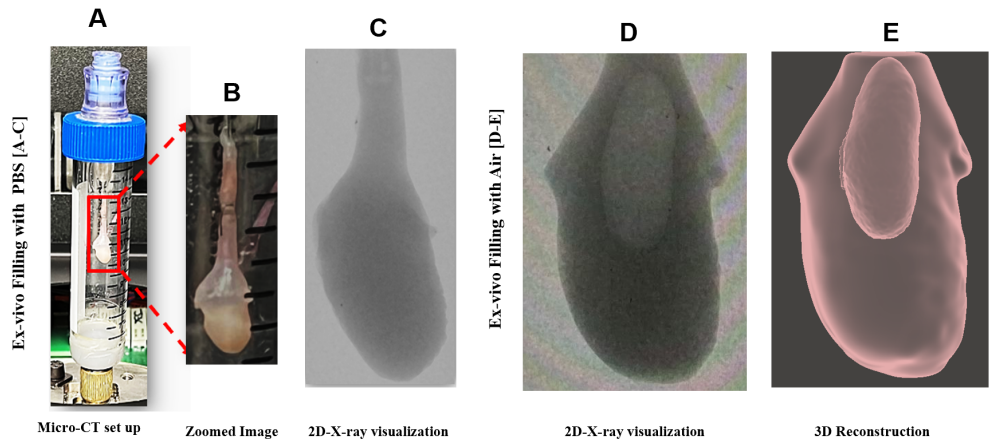

**Figure 1.** (A) In preparation for micro-CT scanning, the catheterized bladder was mounted in a tube and sealed with an airtight Luer lock. In (B), the mounting needle can be seen in the bladder lumen, secured with sutures. While the lumen cannot be identified in the 2D micro-CT data sets when filled with PBS (C), the lumen is clearly distinguished when filled with air (D). The 2D stacks from the air-filled data sets could then be 3D reconstructed (E) to obtain distinct lumen and surrounding bladder wall regions.

**Multiphoton Microscopy (MPM) Imaging of Resected Bladder**

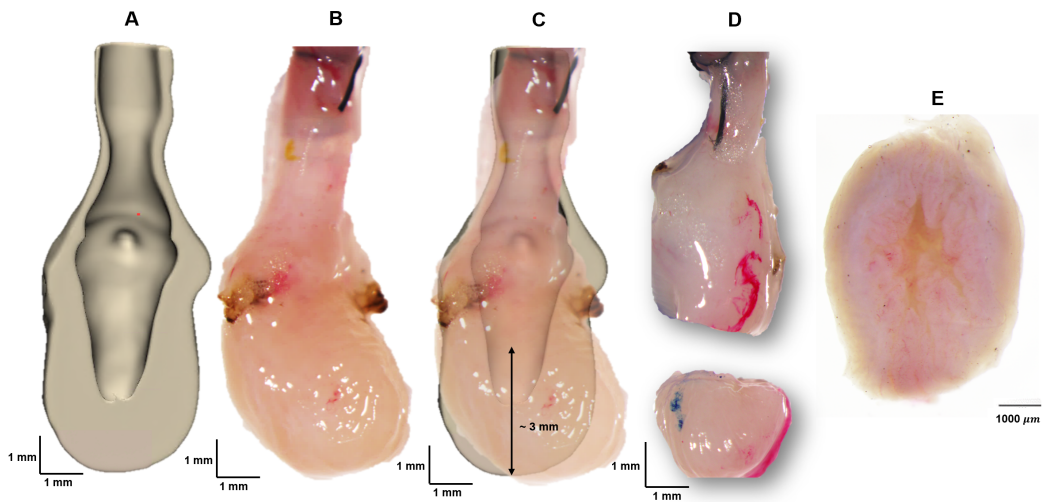

**Figure 2.** A 3D reconstructed micro-CT data set was bisected to display the distinct lumen and albumen regions (A). The dissection scope image of the same fresh bladder (B) was then compared with (A) to identify the location for the transverse cut (C) that would ensure both lumen and surrounding thickened wall would be visible. A dissection scope image looking into the cut dome region (D), image of bladder dome surface from MPM microscope lens prior to scan (E).

To augment the micro-CT analysis, scanning multiphoton imaging was performed on a fresh (unfixed) voided female SD rat bladder looking down into a portion of the dome, Fig. 2. Briefly, micro-CT analysis was first performed on a voided, air-filled bladder (as above) to determine the location of open lumen within the bladder. Based on the 3D reconstructed bladder images, a transverse cut was made 3 mm from the base of the bladder wall. The bladder dome was subsequently immersed in Hank's buffer solution, and scanning multiphoton microscopy was performed directed down into the lumen (Nikon A1R MP HD, Tokyo, Japan).

A Nikon A1R MP HD multiphoton microscope, equipped with the Nikon Ni-E upright motorized system and Chameleon Laser vision, was utilized for extensive area scans of bladder specimens. Additional high-resolution 500 μm x 500 μm scans of sectioned bladder tissue were performed (with the scanning feature disabled). The imaging employed an APO LWD 25x water immersion objective lens with a numerical aperture of 1.10. The laser emission wavelength was set to 830 nm, and the excitation frequencies were configured to 400 to 492 nm for channel 1 and 500 to 550 nm for channel 2. For scanning, the MPM protocol used the resonant scanning mode and large image scanning function as in[24,25]. The acquired images were reconstructed using IMARIS 9.5.0 (BitPlane AG, Zurich, Switzerland), enabling the extraction of orthogonal virtual slices in the XZ and YZ planes from the 3D reconstructions. In this work, virtual slice thickness was set to 10 μm.

## Ex Vivo Bladder Filling with Contrast Liquid
### Pressure-Volume Measurements from Bladder Filling Experiments

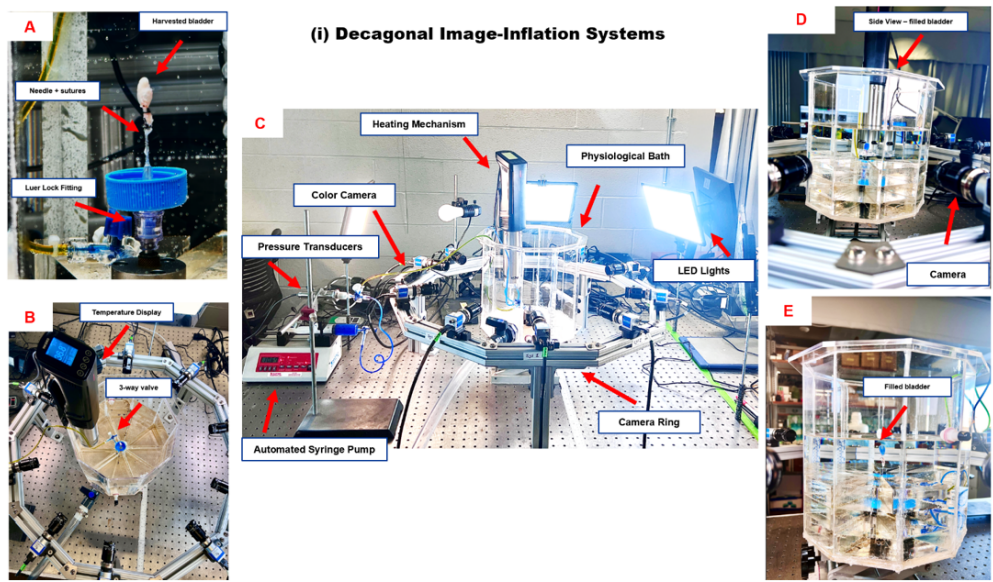

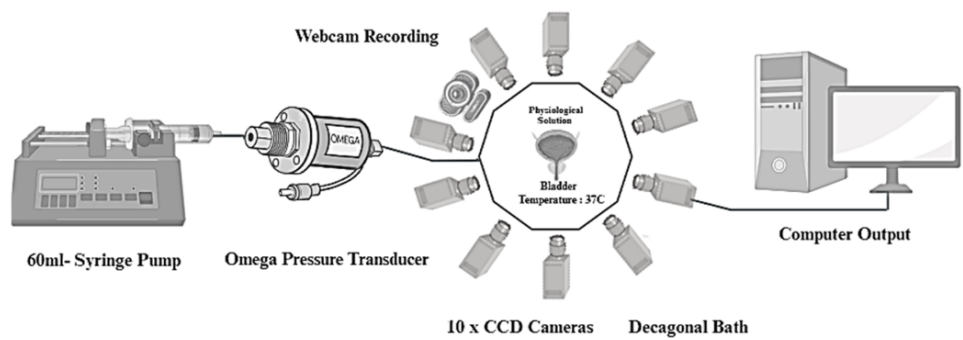

**Figure 3.** Decagonal Image-Inflation Systems(i), [A-E], Schematic of Decagonal Image-Inflation Systems(ii)

Three male bladders were harvested in the voided state following the micro-CT protocol, above. First, the bladder was cannulated and urine removed, as above. Then the air was replaced with a solution of PBS and blue colored dye (watercolor, Dr. Ph. Martin's, Bombay India). using the 1 ml syringe. The dye solution was chosen based on a previously published protocol for imaging the lumen of mice bladders[26]. Each sample was then transferred to a decagonal chamber and mounted upright using a female Luer lock (80147, Qosina Corp., North Ronkonkoma, NY, USA) attached to a 25-gauge needle (B2550, Sterile Blunt Needle, Fisher Scientific). The chamber was filled with calcium-free phosphate-buffered saline (PBS) containing NaCl 137 mM, KCl 2.7 mM, $Na_2HPO_4$ 10 mM, $KH_2PO_4$ 1.8 mM, at pH 7.4. This solution was aerated with a 95% $O_2$ and 5% $CO_2$ mixture and maintained at 37,°C using a Sous Vide Machine (Inkbird, China).

The Decagonal Image-Inflation System setup, as shown in Fig. 3(ii), consists of a 60 ml syringe pump connected to an Omega pressure transducer, which infuses the bladder with the solution. The system features ten CCD cameras arranged decagonally, capturing multi-angle, high-resolution images for a comprehensive analysis of bladder biomechanics. The chamber temperature was maintained at 37°C to simulate physiological conditions. An Ultra 4K HD webcam (Logitech Brio, Lausanne, Switzerland) was integrated with ten 2.3 MP CSI-acA cameras for synchronous recording.

Figure 3(i) provides detailed visual references for the various components of the system. Image (A) shows the setup for bladder attachment, utilizing a needle and sutures with a Luer lock fitting for secure infusion. Image (B) highlights the temperature display and a 3-way valve used for controlling the flow and monitoring solution temperature. Image (C) displays the complete setup, which includes the heating mechanism, automated syringe pump, pressure transducers, camera ring, and LED lights designed to optimize imaging conditions. Image (D) provides a side view of the filled bladder positioned within the physiological bath, with a camera placed for detailed imaging and monitoring. Image (E) shows a close-up view of the filled bladder within the transparent bath, ensuring clear visualization for real-time monitoring throughout the experiment.

The bladder was then slowly infused with the same dye-PBS solution using a 60 ml programmable syringe pump (BS-8000, Braintree Scientific Inc.). Infusion was set at a volumetric flow rate of 20 $\mu$l/min, which is 5 times slower than in-vivo cytometry rate of 100 $\mu$l/min[19,27]. Pressure was recorded using a high-accuracy pressure transducer (PX409, Omega Engineering Inc.) at a 10 Hz sampling rate. Infusion ceased when a maximum pressure of 15 mmHg was reached, visually confirming a full bladder cytometry shape akin to in-vivo ultrasound observations[28]. As in[14], preconditioning was omitted so filling would start from the natural voided state, Figure 3.

*Analysis of Pressure-Volume Data*

Figure 4 displays a typical dataset for pressure versus infused volume during a bladder filling experiment. In all cases, as fluid was slowly infused, there was an initial spike in pressure (highlighted in yellow) followed by a region with gradually increasing pressure (high compliance), followed by a region of diminished compliance. This pressure spike has been previously reported in inflation studies as well as in rat cystometry studies[14,29]. As in prior work[14], pressure-volume data was analyzed for filling studies after this initial spike.

Three regimes were identified in the pressure-volume data: toe, transition and high pressure regimes. As a central objective was to assess the compliance in these regimes, the data were first converted to volume-pressure data. To determine the volume defining the end of the toe regime (beginning of the transition regime), denoted by ($V_{t1}$), a linearly fit was iteratively applied to an increasing range of data until the data deviated from the linear fit beyond a threshold value. In particular, $V_{t1}$ was defined as the volume at which the error in the linear fit for volume was less than 2.5% of the maximum volume for the fitted range, Fig 4. The corresponding pressure is denoted by $P_{t1}$. The volume at the end of the transition regime (and beginning of the high pressure regime) is denoted as $V_{t2}$. Here $V_{t2}$ was similarly defined relative to a linear fit to the high pressure data. In this case, a choice had to be made about the highest volume data for the linear fit. The transition regime was defined as the intermediate regime, $V \in [V_{t1}, V_{t2}]$ with a corresponding pressure range of $[P_{t1}, P_{t2}]$.

The compliance of the toe and high pressure regimes were then calculated as the slopes of these respective lines. The linear range and corresponding $R^2$ value were determined for both the toe and high pressure regions.

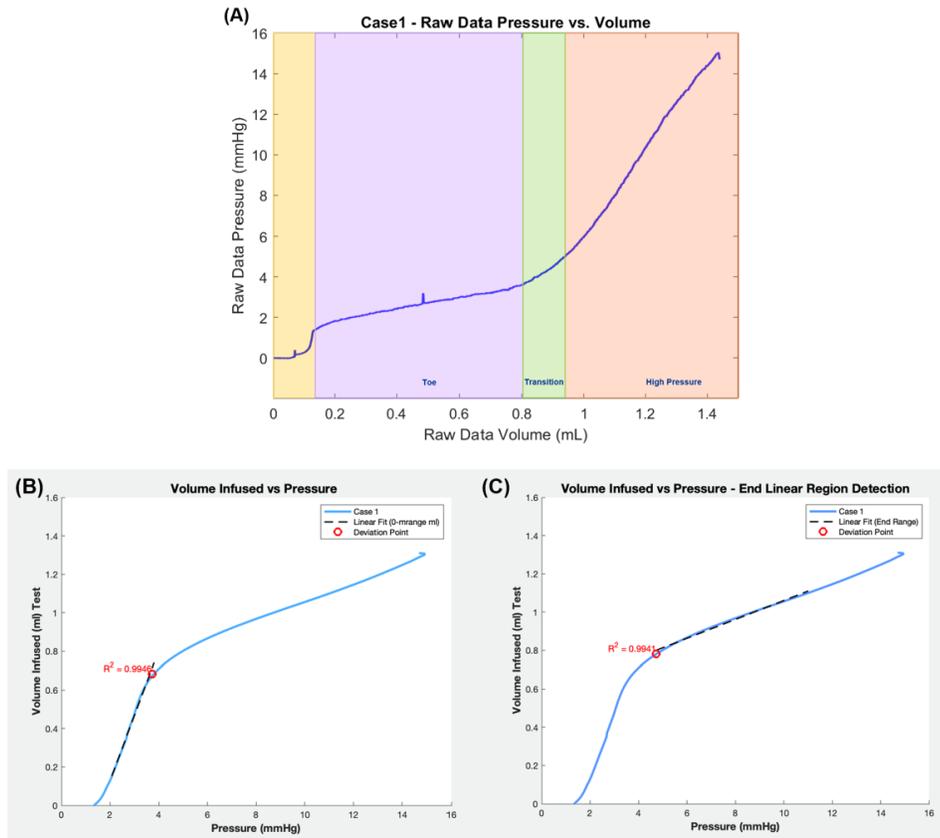

**Figure 4.** Analysis of bladder compliance for illustrative Case 1. (A) shows the raw data pressure vs. volume curve with three filling regimes (Toe, Transition, High Pressure). The linear fits used to identify the beginning and end of the transition regime are shown in (B) and (C) respectively. Here the infused volume and filling pressure are shown, adjusted to remove the initial pressure spike (B, C). For (B), a linear fit ($R^2 = 0.9946$) is applied from 0.15 to 0.68 ml with transition values of $V_{t1} = 0.68$ ml and $P_{t1} = 3.72$ mmHg. For (C), a linear fit to the high pressure regime is shown ($R^2 = 0.9941$) for across the volume 0.78 to 1.00 ml, marking $V_{t2} = 0.78$ ml and $P_{t2} = 4.73$ mmHg. These corresponding compliances in the toe and high-pressure regions are 0.35 ml/mmHg and 0.050 ml/mmHg respectively.

## Results

### Differences in bladder morphology between voided and filled state

Three bladders were imaged in the voided and air-filled state using high resolution micro-CT, without preconditioning. The 3D reconstructed micro-CT data sets for the bladder lumen and ablumen along with corresponding wall thickness contours show marked differences in lumen shape between the voided and filled states, Fig. 5. In the voided bladder, the lumen extends only partially down the length of the bladder and the bladder appears to be fully closed in the apex of the dome region, Fig. 5 (a, i, q). As a result, the wall thickness in the voided bladder is highly heterogeneous, being substantially thicker at the apex. Across all three bladders, the average wall thickness in the voided state is 670 microns with an interquartile range (IQR) of 720 microns, Table 2. This heterogeneity is apparent in the sectioned view of the bladder, Fig. 5 (b, j, r), and quantified in color contours of wall thickness. The average median thickness in the dome was 2.40 mm ± 0.37 mm, while the median thickness in the mid-bladder and trigonal areas was 0.67 mm ± 0.11 mm, Fig. 5 , (b-d, j-l, r-t). Conversely, in the distended state all three bladders demonstrated a relatively uniform and thin wall, Fig. 6, with an average median thickness of 0.10 mm ± 0.02 mm and IQR of 40 microns (Table 2).

The volume of the three bladders showed a marked increase from the voided to filled state, averaging 62-fold, Table 1. Specifically, Bladder A exhibited a 16-fold increase, Bladder B a 92-fold increase, and Bladder C an 78-fold increase (Table 1). Due to the substantial increase in size, the filled and voided bladders in Fig. 5 and 6 are drawn to different scales.

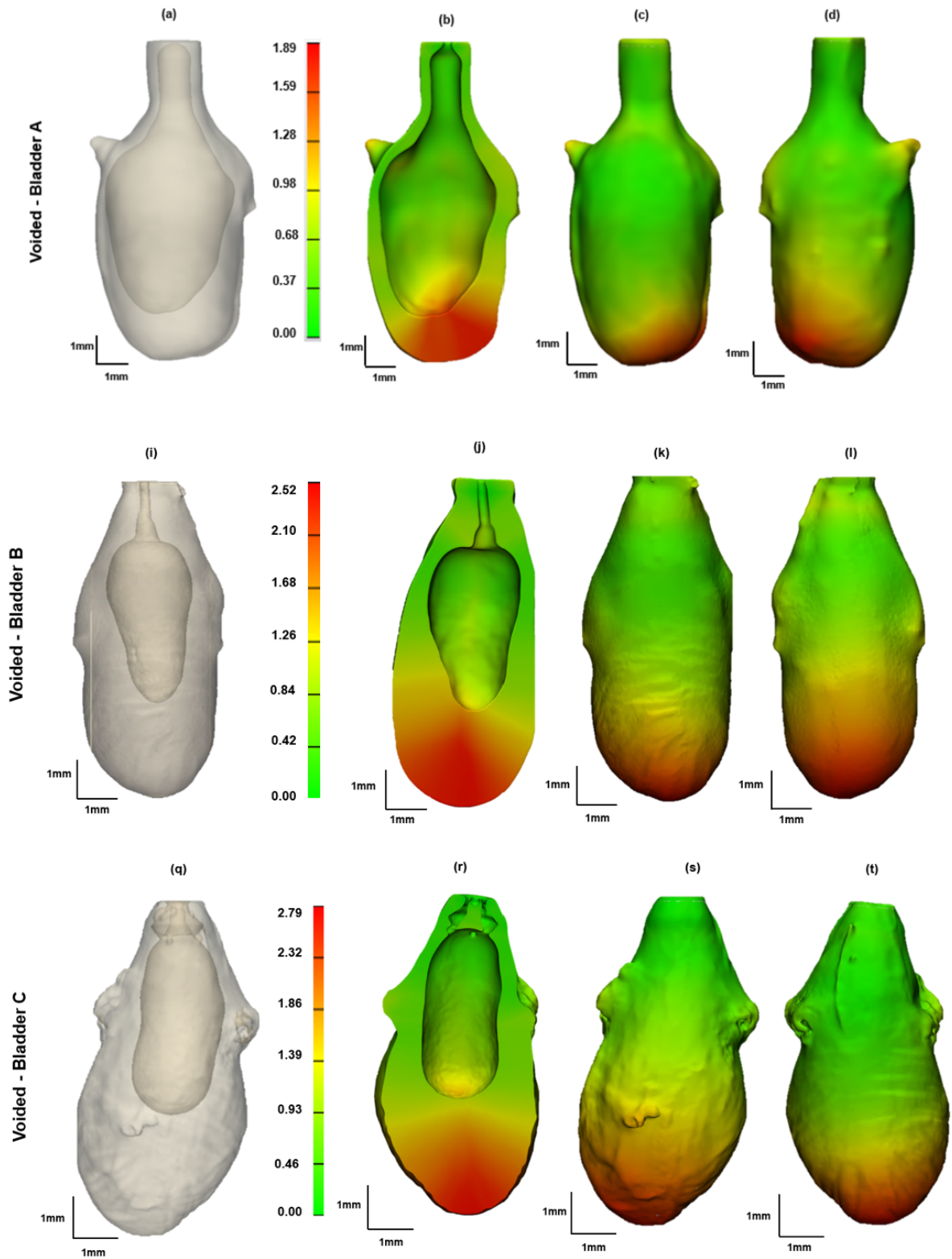

**Figure 5.** Bladder wall thickness in voided inflation state, N=3. Column 1 is the transparent view of the voided volumes and column 2 shows a cross-section of the 3D reconstructed micro-CT data. Columns 3-4 show color contours for wall thickness. All measurements are in millimeters (mm).

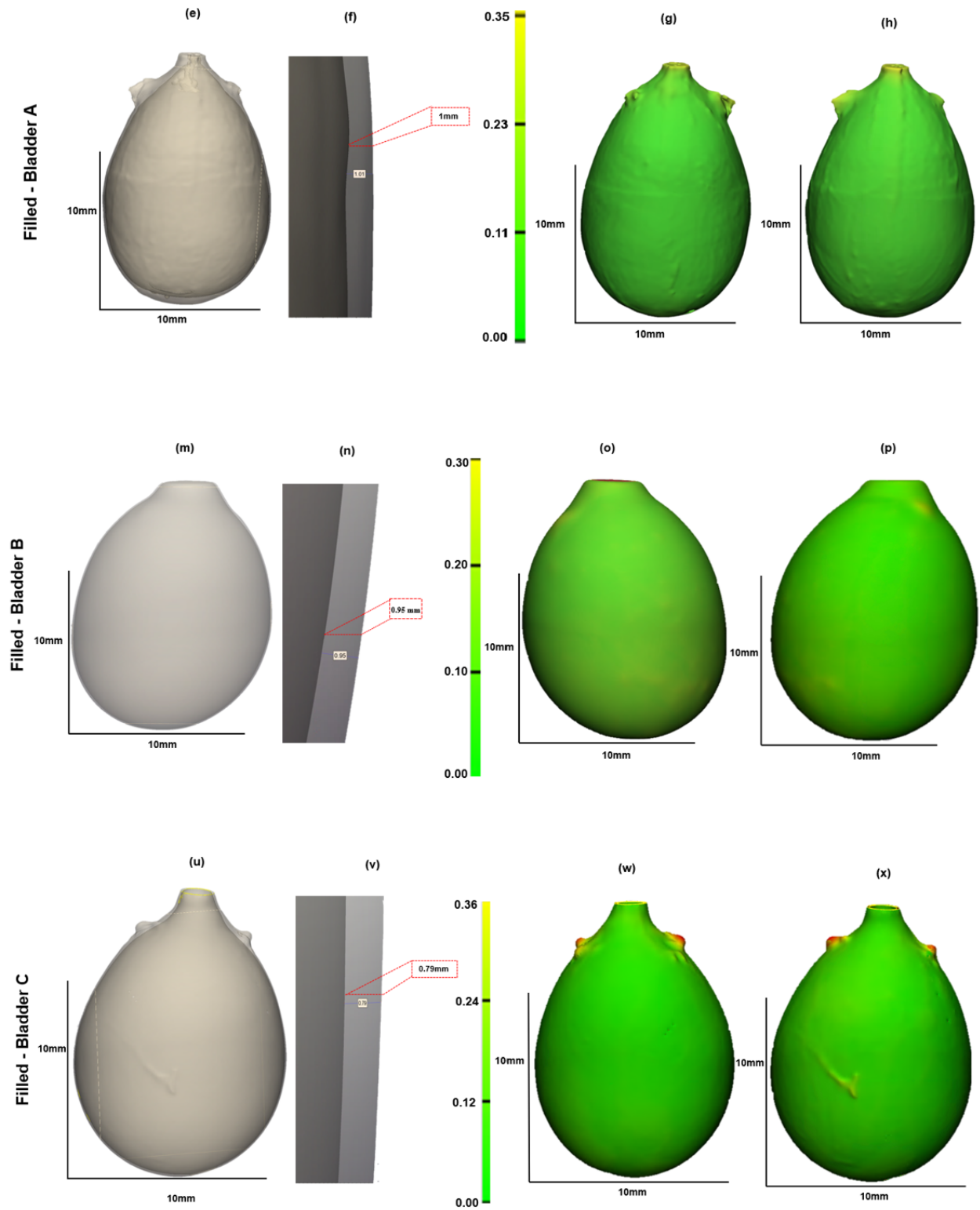

**Figure 6.** Bladder wall thickness at filled inflation state, N=3. Columns 1 shows a transparent section of the 3D reconstructed micro-CT data with a zoomed-in section of the lumen and wall. Columns 3-4 display color contours for wall thickness. All measurements are in millimeters (mm).

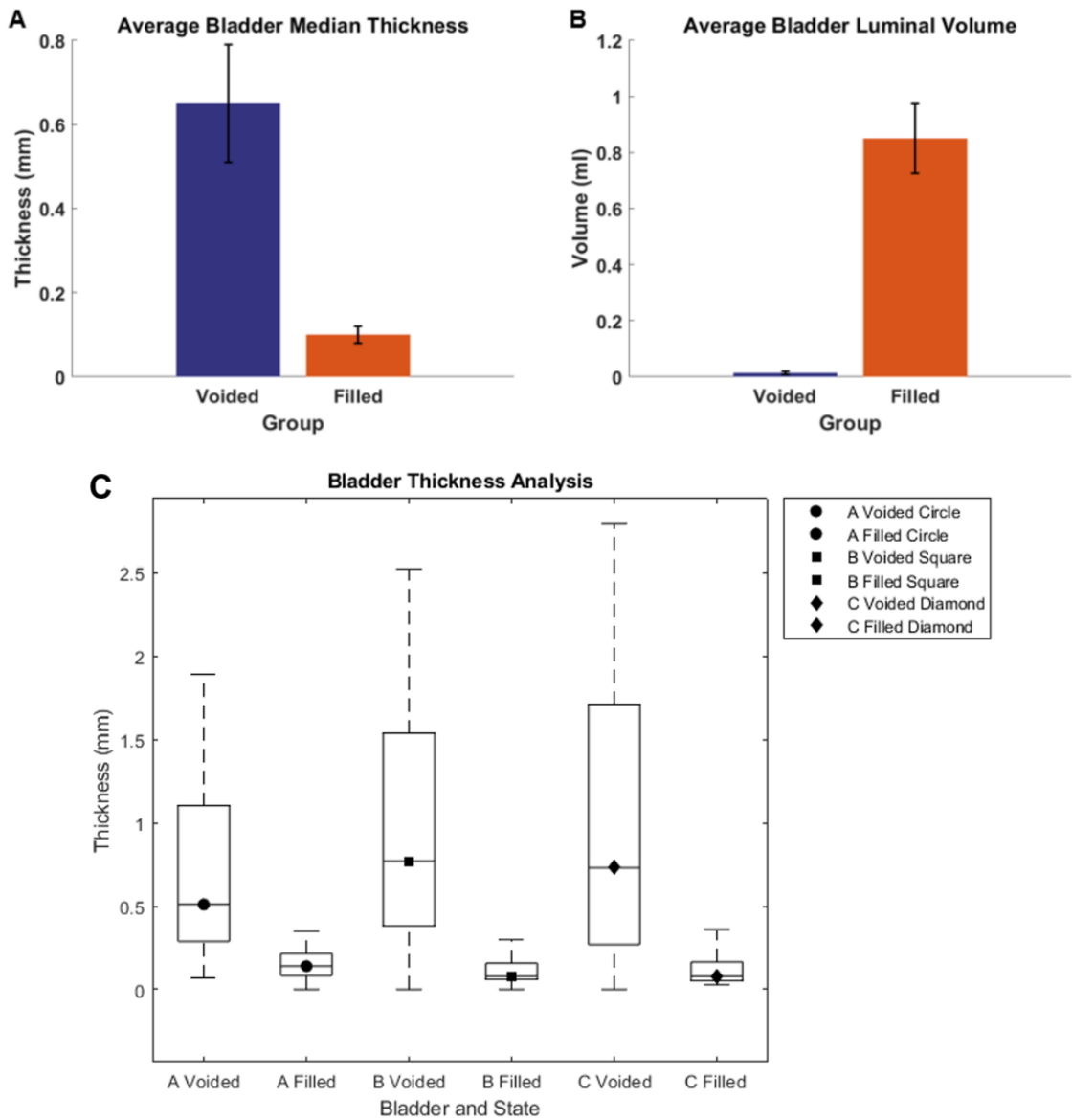

**Figure 7.** (A) average of bladder median thickness. (B) Average of bladder luminal volume for voided and filled state. (C) Box plot of wall thickness in voided and filled state, N=3.

| Bladder | Residual Volume (ml) | Filled Volume (ml) | Ejection Fraction (%) |
|---|---|---|---|
| A | 0.051 | 0.808 | 93.70 |
| B | 0.011 | 1.008 | 98.92 |
| C | 0.009 | 0.706 | 98.77 |
| Average | 0.023 | 0.841 | 97.13 |
| STD | 0.019 | 0.125 | 2.42 |

**Table 1.** Bladder volumes in voided and filled states calculated in 3D reconstructed volumes generated from micro-CT data sets.

| Bladder | State  | Q1 (25th) | Median | Q3 (75th) | Maximum | IQR  |
|---------|--------|-----------|--------|-----------|---------|------|
| A       | Voided | 0.36      | 0.51   | 0.84      | 1.89    | 0.48 |
| A       | Filled | 0.11      | 0.14   | 0.17      | 0.35    | 0.06 |
| B       | Voided | 0.51      | 0.77   | 1.21      | 2.52    | 0.70 |
| B       | Filled | 0.08      | 0.08   | 0.11      | 0.3     | 0.03 |
| C       | Voided | 0.36      | 0.73   | 1.35      | 2.79    | 0.99 |
| C       | Filled | 0.06      | 0.08   | 0.10      | 0.36    | 0.04 |

**Table 2.** Bladder thickness analysis in voided and filled states (mm).

## Mechanism for low residual volume in voided state-MPM imaging

To further investigate the closed dome seen in the micro-CT data sets, Fig. 5, the interior of the voided bladder dome, Fig. 2, was imaged with high-resolution scanning multiphoton microscopy (MPM), Fig. 8.

The internal morphology of the bladder wall was visible under MPM without staining or fixation, due to the second harmonic signal of collagen fibers (red), Fig. 8. A system of large folds can be seen across the entire surface, Fig. 8 (A). In the zoomed image (B), the large folds, with a width on the order of 750 microns, can be seen to be formed from multiple subfolds, Fig. 8 (C, D). The folds do not extend across the entire wall thickness, but rather appear to be formed from the inner layer alone, Fig. 8 (E).

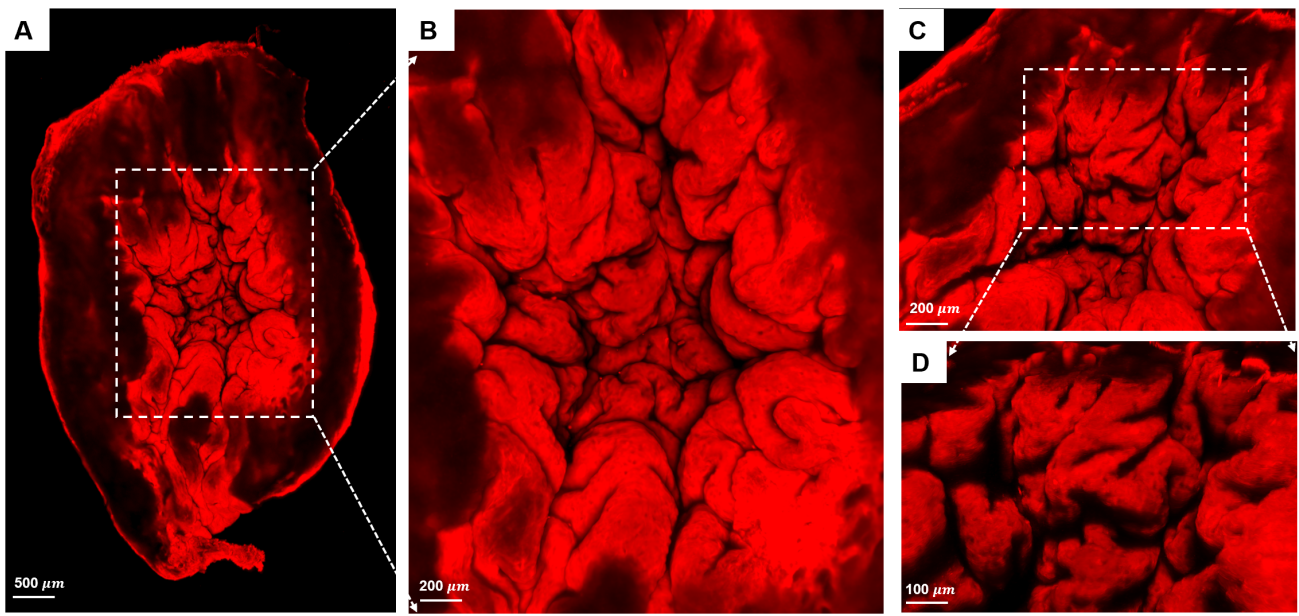

**Figure 8.** Scanning MPM images of internal surface of bladder dome. (A) Overview of the entire lumen of the bladder dome. The thickened wall can be seen to contain large-scale folds that are more clearly visible in zoomed views (B-D). The folds have a hierarchical folding pattern with the larger folds being composed of smaller subfolds (C, D).

## Pressure vs Volume relationship during bladder filling

Data for pressure versus infused volume for three cases are shown in Fig. 9. Two of the curves show an initial region of high compliance followed by a transition to a region of much lower compliance. For Case 3, the pressure volume data also starts with high compliance. However, unlike the other cases, the transition region is followed by a second region of relatively high compliance.

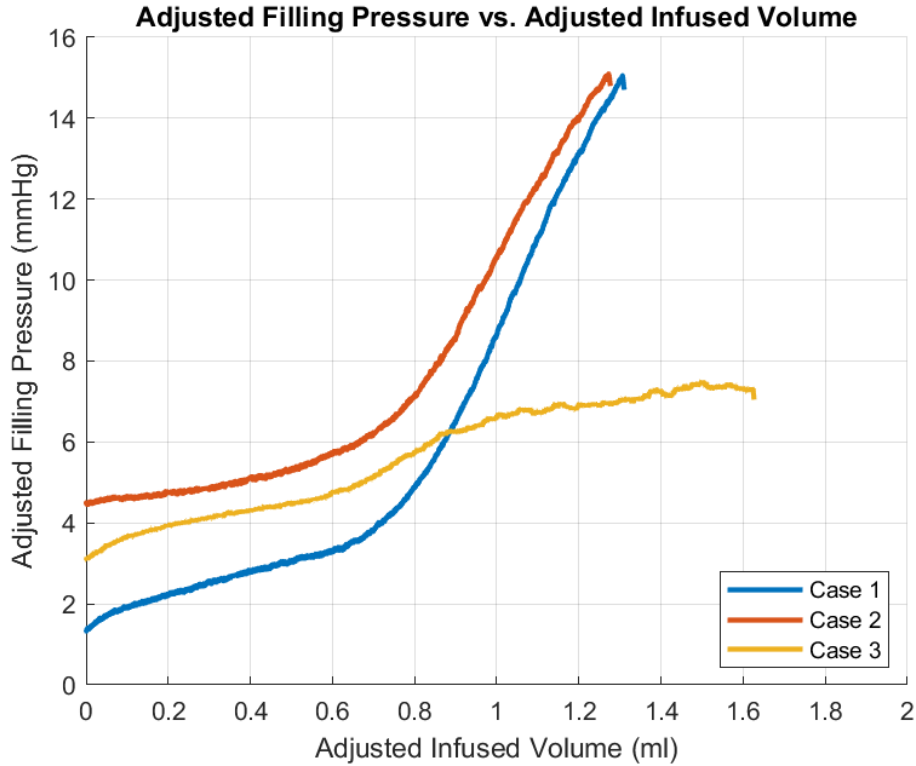

**Figure 9.** Pressure vs infused volume during bladder filling.

The methodology for calculating the beginning and end of the transition regime is illustrated for Case 1 and shown in Fig. 4. Values of pressure and volume at the beginning and end of the transition regime for each case are given in Table 3.

### Relationship between bladder morphology and high compliance

Images of the bladder during the infusion process are provided Fig. 10 displayed for 0.1 ml increments. The images are separated into toe, transition and high pressure regimes using results in Table 3. Two distinct regions can be seen within each dye-filled bladder. A blue region, where the dye-filled lumen can be seen through the wall of the bladder. Additionally, "white" areas are visible and correspond to regions where the bladder is fully closed (the lumen is absent) or the wall is sufficiently thick that the wall is opaque. Initially, this white region extends well into the mid-dome region. As filling increases, the white region progressively diminishes until there is only a thin sliver left, Fig. 10.

| Measured Variable | Case 1 | Case 2 | Case 3 |
|---|---|---|---|
| $V_{t1}$ (ml) | 0.68 | 0.44 | 0.63 |
| $V_{t2}$ (ml) | 0.78 | 0.77 | 0.85 |
| $P_{t1}$ (mmHg) | 3.72 | 5.10 | 4.80 |
| $P_{t2}$ (mmHg) | 4.73 | 6.81 | 6.02 |
| Compliance in toe region (ml/mmHg) | 0.35 | 0.57 | 0.52 |
| Compliance in high pressure regime (ml/mmHg) | 0.05 | 0.06 | 0.86 |

**Table 3.** Volume and pressure at interface of toe and transition regimes ($V_{t1}, P_{t1}$) and interface of transition and high pressure regimes ($V_{t2}, P_{t2}$) along with bladder compliance in toe and high pressure regimes for Case 1-3.

In the toe region, the bladder accommodates large changes in volume with smaller increases in pressure. Two mechanisms for this high compliance can be discerned from Fig. 10. In the early stages of the toe regime, volume increase appears to be due to progressive unfolding of the dome, seen by the diminishing white region in the dye-filled bladders coupled with changes in width and length of the bladder. With increased filling in the toe regime, the bladder width and height increase markedly with negligible change in size of the folded region. In the high pressure regime, the width and height show little change despite continued increase in volume, and the size of the white region appears unchanged. The increase in intensity in the blue dye color in the transition and high pressure regime compared with the toe regime suggests the wall has also thinned.

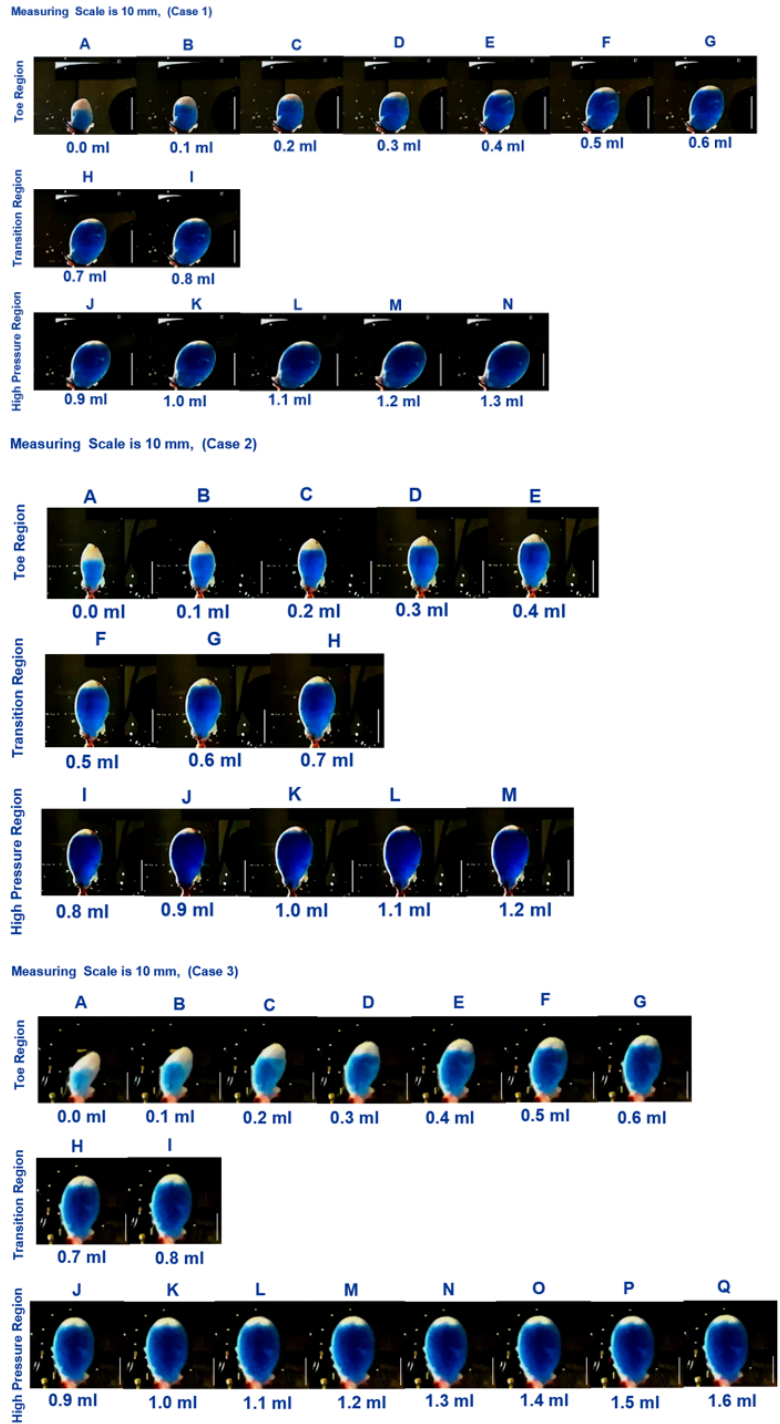

**Figure 10.** Images of the bladder during infusion with blue dye solution. The blue dye can be seen in open regions of bladder. Images are provided in 0.1 ml increments and separated into toe, transition and high pressure regimes. Scale bar is 10 mm (Case 1, 2, 3).

## Discussion

The urinary bladder, a compliant reservoir within the lower urinary tract, is designed to accommodate significant volumes of urine with minimal increases in transmural pressure, a property critical for maintaining urinary tract homeostasis and protecting upper tract function[7]. Numerous prior works have conjectured that small undulations (rugae) in the inner wall are responsible for the high compliance of the healthy bladder[19,30,31]. However, this has not been directly proven and has remained an open question. In this work, we re-investigated this open question, leveraging advances in bio-imaging technology. Combining data from three imaging modalities, we discovered that folds in the inner wall, an order of magnitude larger than previously reported rugae, play an important role in early filling compliance.

A second remarkable feature of the bladder is its high voiding efficiency of over 90%. Remarkably, from our in vitro studies, the rat bladder exhibits a voiding efficiency of approximately 90–95%. Notably, Cheng et al.'s in vivo study of voiding in healthy rats (n =12) using urodynamics, also found a high voiding efficiency of over 95 %. The high efficiency of voiding in the bladder is particularly remarkable when compared with the corresponding values for the heart (ejection fraction). For example, the heart ejection fraction ranges from 52–72% for men and 54–74% for women[32]. The heart ejection fraction for healthy rats are similarly lower, with reported values of $69 \pm 3\%$[33]. This comparison underscores the bladder's high functional efficiency, ensuring nearly complete voiding of its contents while maintaining low pressures during the initial filling phase (toe region), contributing to its critical role in urinary system health. As elaborated on below, our results suggest the large scale folds in the inner wall not only play an important role in the bladder's high compliance, but also in its high voiding fraction.

It is interesting to reflect on the physiological demands that have led to these vastly different ejection/voiding fractions. In the case of the bladder, elevated residual volume is associated with urinary tract infection due to growth of bacteria within the bladder. For example, in a study of healthy men (average age 62), Truzzi et al.[34] determined that a post-void residual volume exceeding 180 ml significantly increases the risk of bacteriuria. This is a problem unique to the bladder, related to the composition of urine, the long bladder retention times, which can be on the order of 8 hours during the sleep cycle, and the the ability for bacteria to to gain access to the bladder through the urethra.

### *Bladder contraction causes buckling of the inner bladder wall*

One of the remarkable features of the healthy human male bladder is its ability to increase in volume by 28.4-fold (mean age: 24 years, $n = 28$), filling from a contracted mean (residual) volume of 19.7 mL to include an addition 541.3 mL (voided volume), as demonstrated in pressure-flow studies conducted on normal adult volunteers[18]. In the case of the rat bladder studied here, the volume increased by an average of 62-fold. Such large scale increases necessitate significant areal changes across all layers of the bladder wall. In the detrusor smooth muscle (DSM) layer, contraction and relaxation of smooth muscle during voiding and filling contribute to the areal changes, though there is a need for futher investigation of this process on a quantitative level. In contrast, the lamina propria layer lacks mechanisms for active contraction and relaxation, requiring alternative methods to accommodate large deformations. A key finding of this study is that the DSM contraction in the voided state is accompanied by large-scale buckling of the lamina propria and urothelium layers. During filling, the flattening of the folds facilitates the large increase in volume. Notably, the amplitude of these folds is approximately ten times larger than the amplitude of previously reported rugae, which have been measured at approximately 50-100 μm in height[19,30,31].

### *Contribution of folds to high voiding fraction*

A second key finding of the present work is the role of these large scale folds in achieving the remarkable voiding fraction of the bladder - greater than 94%, Table 1. Specifically, we demonstrated that effective voiding is made possible by large folds in the inner layer of the bladder dome. These folds effectively fill the lumen of the contracted dome, driving the urine toward the trigonal area. This conclusion is based on results from two complementary imaging modalities. First, 3D reconstructions from high-resolution micro-CT data consistently showed voided fractions above 94%, Fig. 5. These micro-CT data sets also revealed the voiding process is not simply a uniform contraction of the filled bladder. If it were, a uniformly thin walled filled bladder would contract into a smaller shape while maintaining uniform wall thickness. Instead, the voided bladder exhibited highly heterogeneous wall thickness, with the lumen of the upper dome appearing fully closed and wall thickness gradually decreasing toward the trigonal region, Fig. 5. Second, complementary multiphoton imaging of the upper dome region revealed that the lumen, which appears closed in the micro-CT images, Fig. 5, is, in fact, filled with a densely folded layer formed within the contracted detrusor muscle layer Fig. 8. In the dome apex, these folds effectively fill the lumen, driving urine toward the trigonal area and preventing urine from being trapped in the upper dome during voiding.

The high voiding fraction measured ex-vivo in our study using micro-CT aligns with in vivo measurements obtained from awake cystometry. In particular, for the current ex-vivo study of bladders from 4 month Sprague Dawley male rats, the mean filled volume was 0.841 ± 0.125 ml with an average residual volume of 0.023 ± 0.019 ml, Table 1. This corresponds to an average voided fraction of 97± 3%. Previously reported data from in vivo bladder cystometry for 4 month male Sprague Dawley rats (N=12) revealed an average voided volume of 0.84 ml ± 0.03, with an average residual volume of 0.02 ± 0.008 ml[27], corresponding to a filled volume of approximately 0.86 ml and a voiding fraction surpassing 95%. Moreover, this high voiding fraction is also a feature of healthy voiding in humans. For example in a study of healthy males from Wyndaele in 1999, bladder capacities measured using urodynamic study had mean voiding fraction of 96.49% for men[18].

### *Role of folds in bladder compliance*

We identified three distinct regions in our pressure-volume curves: the toe, transition, and high-pressure regions (Fig. 10). A third key finding in the work is that the high compliance of the bladder in the first half of the toe region corresponds to the opening/flattening of the folds as the lumen increases in size. This result was proven by imaging voided bladders during the filling process while simultaneously measuring transmural pressure and infused volume, Fig. 10, Fig. 4 and Table 3. In these studies, folded regions in the initial voided bladder dome exclude dye and appear as opaque caps at the top of the dome, Fig. 10. The unfolding process could be seen as an increase in the size of the dye filled region and decrease in the size of the opaque cap. In all cases, as the bladder was inflated through the toe region (high compliance region), the folded region gradually opened, Fig. 10. In the remaining toe regime, the increase in volume was accompanied by substantial increase in the size (width and height) of the lumen. In the transition region and high stiffness regions, the infused volume continued to increase with little or no apparent change in the size of the opaque cap and a negligble change in width and height of the bladder.

### *Mechanisms for bladder compliance beyond unfolding*

Cheng et al. evaluated the specific contributions of the different wall layers to bladder compliance in adult (12 months) and aged (21-24 months) rats using strain controlled biaxial testing under multiphoton microscope. While in the present study the starting point for mechanically testing was the voided bladder, in this earlier study a square specimen was cut from an opened unloaded bladder, that was not contracted (voided). In these unloaded specimens, the inner wall of the unloaded bladder had undulations on the order of 50-100 microns, corresponding to the dimensions of the rugae in prior reports,[19,27]. Using MPM imaging simultaneous with biaxial testing, it was discovered that rugae in the LP combined with tortuosity of collagen fibers in the DSM provide mechanisms for compliance in non-voided bladder. Namely, the rugae flattened without any discernible recruitment of collagen fibers in the high compliance (toe) region of their study. As the loading extended into the transition regime of that study, a gradual coordinated recruitment of collagen fibers between the LP and DSM layers was seen, corresponding to a transition to the high stiffness regime.

In relating the findings of Cheng et al. to the current study, we conjecture the compliance mechanisms identified by Cheng et al. are activated once the large scale folds are flattened during bladder filling. We note the square specimens used in[19] were taken from a non-voided bladder in the mid region of the dome. The unloaded wall thickness for the square specimens was 0.56 ± 0.09 mm. We can see from the micro-CT data that the unloaded trigonal region of the voided bladder, Fig. 5, is on the order of 0.5 mm, similar to that of that of this unloaded square specimen.

As expected, since the bladder was not voided in the study of Cheng et al., no large scale folds were seen. In the present study, the unfolding process can be seen to progress from slightly above the trigonal area up towards the dome, Fig. 10. While we do not have quantitative information about the unfolded wall thickness during filling, in the filled state, it can be seen to be relatively uniform across most of the bladder wall, including the trigonal area, Fig. 10. Moreover, in the toe region and into the high stress regime of the filling bladder, Fig.10, there is relatively little change in size of the opaque region. Nonetheless, the volume increased by an average of 30% in the high pressure regime, suggesting a second mechanisms for compliance. We conjecture the mechanisms reported by Cheng et al., provide a second source of compliance, activated in regions in the bladder wall where the large scale folds are flattened, possibly explaining the high compliance in the latter part of the toe regime. This conjecture will be investigated in a future study. As elaborated on below an additional contributing factor to the high pressure compliance is due to the larger volume at high pressures. Namely, as the bladder enlarges, proportionally less strain is required for a given increase in volume.

### *Prior compliance studies in whole bladder filling*

Most prior investigations of whole bladder filling did not study the voided bladder. These studies started with a bladder state when the large scale folds seen here were absent, either because the bladder was harvested in a non-contracted (voided) state or due to preconditioning[10]. The motivation for preconditioning soft tissues before mechanical testing was discussed as early as 1967 by Fung[35–37] who noted that preconditioning induces a stable, "nonlinear, pseudo-elastic" state, needed to achieve a "steady state" for precise mechanical assessment. In the case of bladder, the preconditioning process will stretch out the contracted DSM and unfold the inner wall layers, shifting the bladder out of the toe region. Hence, preconditioning is inconsistent with the objectives of the present work.

A few prior studies omitted preconditioning in their studies of bladder filling[13,14]. In an important investigation by Parekh et al.,[13], the bladder was not preconditioned, however, it was partially perfused before the filling experiments were initiated so that strain markers could more effectively be applied to its exterior surface. In particular, the bladder was infused with 0.3 ml liquid, shifting the subsequent infusion experiments to a regime corresponding to the middle to end of the toe region in our study. Parekh et al. also reported an initial high compliance response over a pressure range of 2-3 mm Hg. Neither the source of this high compliance nor heterogeneity of wall thickness across the bladder were the focus of this work and were not investigated.

The current clinical definition of compliance is the inverse of the slope of the line joining two points on a highly nonlinear curve. As these two points bridge the transition regime, they will be highly sensitive to numerous factors such as the length of the toe and transition regions. By defining three distinct regions in our pressure-volume curves, we are able to provide two distinct compliance values. This approach also provided results for the total infused volume at the end of the toe regime and the total infused volume at the onset of the high pressure (low compliance) regime. It is clear that the current clinical definition of compliance will not be able to distinguish between these features of bladder response.

### *Application of Idealized Geometry in Bladder Outlet Obstruction (BOO)*

Historically, numerous models, including our own, have idealized the bladder as a spherical or ellipsoidal structure with a homogeneous wall thickness ,[10,14,38]. This simplification facilitates analytical solutions and interpretation of mechanical data. These solutions are particularly important in computationally intensive studies such as bladder remodeling. For instance, a recent study on the growth and remodeling of bladders with BOO,[27], utilized the filled (spherical) bladder as a reference configuration to define the homeostatic stretch of collagen fibers and smooth muscle cells. Our current findings suggest these idealizations are appropriate for studies of the bladder in the high pressure regime.

However, the present analysis of the voided and filled states, Fig. 7, revealed that even in healthy bladders, the voiding process induces spatial variations in wall thickness and shape, which are not adequately represented by a simple constant thickness spherical model. Therefore, to understand bladder dysfunction during voiding and filling, it will be essential to understand how these changes to the folding and filling process are altered across the bladder wall. Moreover, it is also possible that even a fully distended bladder may exhibit heterogeneity in wall thickness due to pathological conditions and aging[22].

These results also underscore the value of advanced imaging modalities, such as high-resolution micro-CT that are needed to accurately delineate bladder morphology and identify region-specific alterations in wall thickness throughout the filling and voiding cycles. Such high-resolution data are indispensable for the development of computational models of bladder function such as those used to investigate the evolving functionality of the bladder in conditions like BOO.

## Study Limitation & Future Directions

While this study provides insights into the complex filling process there are remaining open questions. One question relates to the shape of the pressure-volume curve in the high pressure regime. While the shape of the toe regime was consistent across all samples, the high pressure regime displayed two qualitatively different shapes. In all cases, the high pressure response was a monotonically increasing function of volume. However, for Cases 1 and 2 the compliance dropped sharply between the toe and high pressure regimes which is more common as noted by the literature[13,14], while that of Case 3 remained relatively high. Both of these qualitative shapes in the bladder pressure volume curves have been previously reported[10].

In searching for a mechanistic explanation for these differences, it is important to recall the pressure-volume curves obtained during bladder inflation studies reflect the response of the whole organ, not just the material properties of the wall layers. This was already made clear for the high compliance regime where, folding of the inner wall layers played an important role. There are numerous factors influencing the bladder compliance in the high pressure regime, some of which are competing effects. In particular, while the recruitment of collagen fibers during inflation will tend to decrease compliance, the effects of increasing lumen size with inflation will act in the opposite direction. This latter effect can be seen from even the simple kinematics of inflation of a spherical membrane, where the change in volume depends on the cube of stretch,

$$\frac{\Delta V}{V_0} = \lambda^3 - 1 \tag{1}$$

Here, $\lambda$ is the stretch, equal to the current radius divided by the reference radius and $V_0$ is the corresponding reference volume for the bladder lumen, $\Delta V$ is the difference between the current and reference lumen volume. If, as expected, the pressure is a monotonically increasing function of stretch, then the compliance will increase faster than $\lambda^2$. Relating this to the curves in Fig. 9, we conjecture the walls in Case 1 and 2 stiffen fast enough to overcome the competing effect of increasing volume. In contrast, for Case 3 bladders, the opposite effect is seen. This conjecture will be investigated further in future studies.

Another unexplained feature of the inflation curve is the steep rise and oscillation in pressure at the onset of filling, despite the use of calcium channel blockers, Fig. 4. This feature of the pressure volume curve has been previously reported by Trostorf et al. in an ex vivo study of passive bladder inflation in pigs[14]. In their study, an initial spike was observed in all six inflation datasets, with a spike of between 1.8 and 3.7 mm Hg (Fig. 3 of[14]). Importantly, an initial pressure rise was also seen in vivo during rat cystometry studies[29]. This spike was not reported in other ex vivo whole bladder inflation studies, possibly due to the preconditioning protocol used in these studies,[10].

In the future, it would be of interest to consider the effects of gender and age on the unfolding process and quantitative aspects of the pressure inflation curves. In the present study, we evaluated male rats in the pressure inflation studies. Differences in voiding pressure flowrate curves between male and female rats have already been explored by Streng et al.[39] who found no overall differences in bladder pressure data between the genders. However, they did not study the filling process. Kories et al.[40] examined muscarinic receptors in males and females, and found the numbers and functions of M2 and M3 muscarinic receptors did not differ between urinary bladders of male and female rats, similar to findings in humans.

In any ex vivo study, there is the question of relevance to the in vivo conditions. A crucial choice in the present study was to consider filling in bladders that were harvested in the voided state. This is expected to be a physiologically relevant baseline, distinguishing the approach in this study from most prior work. Further, temperature was maintained at physiological values and filling rates were chosen to be slow enough to reproduce in vivo filling rates. Nonetheless, the filling process in vivo may be affected by numerous factors that are not represented in our study. For example, the effects of external contacts on shape are not considered. Moreover, the bladder was filled with air in one category of experiments in order to image both the internal and external surfaces of the bladder. At the prescribed filling rates, both gas and liquid will provide nearly uniform loading on the surface of the bladder. As expected, we did not find differences in the shapes of the bladder in the two types of experiments. We were able to compare several of the ex vivo results to published in vivo studies. As noted above, the high voiding fraction reported in vivo of over 95% was consistent with data from our ex-vivo study using micro-CT. In addition, the ex vivo filled volume was within 5% of the reported in vivo value. Future in vivo studies will provide further insights on potential differences and similarities of the ex vivo and in vivo filling processes.

In this study, female rats were used for micro-CT examinations of bladder geometry, as shown in Figs 5, 6, and for complementary multi-photon imaging studies. Male rats, on the other hand, were utilized for the pressure inflation experiments. Pressure inflation studies for the bladder have been conducted in various species and for both male and female animals, (e.g.,[10,14]). No qualitative differences have been noted. In the future, it might be of value to directly compare the quantitative response in male and female response during bladder filling and also in mechanical evaluation of the material properties of the bladder wall. Differences in voiding pressure flowrate curves between male and female rats have already been explored by Streng et al.[39] who found no overall differences in bladder pressure data between the genders. However, they did not study the filling process. Kories et al.[40] examined muscarinic receptors in males and females, and found the numbers and functions of M2 and M3 muscarinic receptors did not differ between urinary bladders of male and female rats.

**Applications of this work for diseases with bladder dysfunction**

Improved understanding of the relationship between the pressure-volume curves and the mechanical, structural and geometric properties of the bladder are essential for interpreting clinical data and developing targeted therapeutic strategies, whether pharmacological or surgical. In this work, we discovered that folds form in the domes of healthy rat bladders during voiding and these folds gradually open during filling- unfolding upward from the trigonal region. We determined that there is a corresponding large increases in volume fraction during this unfolding stage and that this process has a corresponding high compliance. We conjecture, that in pathological conditions such as BOO, interstitial cystitis, or post-radiation therapy, changes in the geometry and mechanical properties of the wall will alter the folding process, possibly compromising aspects of the filling and voiding processes. This is a subject of ongoing work.

We anticipate, the mechanisms associated with the distinct phases of filling (toe and high pressure) could be affected differently in disease or with aging, leading to different impacts on compliance values. Recognizing these differences could be important for developing more nuanced diagnostic and treatment strategies that can better address bladder dysfunctions related to aging and various diseases (e.g.[41]).

Surprisingly, we only found one body of clinical research that considered the importance of considering compliance in multiple regions of the pressure volume curve,[41]. In this earlier study, they referred to the toe an high pressure compliance as initial and terminal compliance. In the Ghoniem et. al 1989 study of children with meningomyelocele, the toe region compliance was found to be a statistically significant metric for distinguishing between patients that did well with treatment from those who did not. The latter group were prone to reflux, renal dysfunction and deterioration of the upper urinary tract.Terminal (high pressure) compliance was relatively constant across both response groups. The method we have introduced here provides an objective means of calculating the high pressure compliance. Given the distinct physical mechanisms responsible for the initial and high pressure compliance and these clinical findings, we recommend that both compliances be used in future studies and analysis of bladder function. We also anticipate the values of pressure and volume at the end of the toe region will be important diagnostically.

## Acknowledgments

The authors wish to extend their appreciation to Amanda Wolf-Johnston (School of Medicine, University of Pittsburgh) for her assistance with the surgical dissections of rat bladders for part of the study. They are also grateful to Ricardo Jose Cardoza for his work developing protocols relevant to this study while he was an M.S. student in the Department of Mechanical Engineering and Materials Science, University of Pittsburgh. The authors express their gratitude to Andrew Holmes (Department of Mechanical Engineering and Materials Science, University of Pittsburgh) for his assistance in fabricating the Decagonal Image-Inflation system.

## Author contributions statement

Fatemeh Azari designed and developed the ex-vivo experimental system and executed the inflation experiments, micro-CT imaging, studies and related data analysis. She harvested the intact bladder specimens from SD rats. Fatemeh Azari and Anne Robertson made substantial contributions to the formulation of core concepts and ideas in this study and interpretation of the findings. Fatemeh Azari wrote the initial draft of the paper and, with Anne Robertson, developed and refined the interpretation of the findings and text. Yasutaka Tobe performed the multiphoton analysis of bladder tissue and contributed to the scientific interpretation of this work. Simon Watkins provided expertise in MPM imaging techniques. Paul Watton contributed insights to the research design and his expertise in interpreting the bladder biomechanics. Lori Birder and Naoki Yoshimura contributed to the experimental design, providing scientific insights on bladder function and the animals for the study. Kanako Matsuoka contributed to the experimental design related to the protocol for bladder harvest and handling. Christopher Hardin contributed to the experimental design and interpretation. All authors reviewed the manuscript.

## Provisions for Data Accessibility

The data utilized and examined in this research are not publicly accessible; however, they may be obtained from the corresponding authors upon reasonable inquiry.

## Funding

The authors acknowledge the financial support provided by NIH grants: R01 AG056944 and R01 DK133434.

## Additional information

**Competing interests** The authors assert that they have no competing interests to disclose.